\documentclass[11pt]{article}
\pdfoutput=1

\usepackage{jheppub}
\usepackage[lofdepth,lotdepth]{subfig} 

\usepackage{import}

\usepackage{bigints}
\usepackage{tikz}
\usetikzlibrary{arrows,chains,matrix,positioning,scopes}

\title{Transmuting CHY formulae}
\author[1]{Max Bollmann,}\emailAdd{max.bollmann@physik.uni-muenchen.de}
\author[1]{Livia Ferro}\emailAdd{livia.ferro@lmu.de}

\affiliation[1]{Arnold--Sommerfeld--Center for Theoretical Physics,\\ Ludwig--Maximilians--Universit\"at, \\ Theresienstra\ss e 37, 80333 M\"unchen, Germany }

\abstract{The various formulations of scattering amplitudes presented in recent years have underlined a hidden unity among very different theories. 
 The KLT and BCJ relations, together with the CHY formulation, connect the S-matrices of a wide range of theories: the  transmutation operators, recently proposed by Cheung, Shen and Wen, provide an account for these similarities.
In this note we use the transmutation operators to link the various CHY integrands at tree-level. Starting from gravity, we generate the integrands for Yang-Mills, biadjoint scalar, Einstein-Maxwell, Yang-Mills scalar, Born-Infeld, Dirac-Born-Infeld, non-linear sigma model and special Galileon  theories, as well as for their extensions. We also commence the study of the CHY-like formulae at loop level.}

\begin{document}
\begin{flushright}
{\small LMU-ASC 55/18}
\end{flushright}
\maketitle

\addtocontents{toc}{\protect\setcounter{tocdepth}{1}}
%%%%%%%%%%%%%%%%%%%%%%%%%%%%%%%%%%%%%%%%%%%%%%%%%%%
%%%%%%%%%%%%%%%%%%%%%%%%%%%%%%%%%%%%%%%%%%%%%%%%%%%

\section{Introduction}
\label{introduction}

The past decades have seen tremendous progress in our understanding of the properties underlying the S-matrices of a wide range of theories. Powerful relations and common structures between scattering amplitudes  of different theories have been discovered. In particular, the Kawai-Lewellen-Tye relations (KLT) \cite{Kawai:1985xq}, the Bern-Carrasco-Johansson (BCJ) double-copy relation \cite{Bern:2008qj} and the Cachazo-He-Yuan (CHY) formulation \cite{Cachazo:2013hca, Cachazo:2013iea, Cachazo:2014xea} have shown a surprising universality underlying amplitude construction. In the latter, only very few building blocks are necessary to construct the integrands of gravity (G), Yang-Mills (YM), biadjoint scalar (BS), Einstein-Maxwell (EM), Yang-Mills scalar (YMS), Born-Infeld (BI), Dirac-Born-Infeld (DBI), non-linear sigma model (NLSM) and special Galileon (SG) theories, as well as their extensions, i.e. more complex theories obtained by adding new fields and interactions to the original ones. 
In this direction,  differential operators have been proposed very recently \cite{Cheung:2017ems} which, starting from tree-level gravity amplitudes, produce the S-matrices of various massless theories in arbitrary dimension. 
 These \emph{transmutation operators} were formulated to act on the explicit expressions for amplitudes, written in terms of Lorentz invariants made from momenta and polarization vectors. It was later shown that the action of at least some of them is equivalent to a particular type of dimensional reduction at the action level \cite{Cheung:2017yef}.

In this paper we prove that the transmutation operators applied to the CHY formulae transform integrands of one theory into another.
This provides a further check of the CHY integrands, some of which were only conjectured using ``squeezing" and ``generalized dimensional reduction". This allows to generate all desired integrands, even for extended theories. 
Most importantly, they could be used at loop level to generate  integrands  starting from gravity. At one-loop, CHY-like formulae were derived from ambitwistor strings for gravity and Yang-Mills \cite{Geyer:2017ela} and from the forward limit for scalar fields \cite{He:2015yua}. The loop integrands of these theories are expressed in terms of $(n+2)$-point tree-level integrands localized on the loop-level scattering equations. This is reminiscent of the Feynman tree theorem and was also explored in terms of operators on the sphere \cite{Roehrig:2017gbt}. Therefore, at least at one-loop, the transmutation operators have a similar action to that  at tree-level and allow to generate all desired integrands. 

The paper is structured as follows: in the next Section we recall some notions about transmutation operators, while in Section \ref{CHY} the CHY formulation of tree-level massless amplitudes is 
reviewed. In Section \ref{transm_CHY} we apply the operators to the CHY integrands. In the last Section we discuss loop amplitudes. Conclusions and outlook complete the paper.

\vspace{0.3cm}
\noindent
{\bf Note Added:} After completion of this work, we have become aware of \cite{Zhou:2018wvn}, where similar computations for the CHY integrands are performed.

%%%%%%%%%%%%%%%%%%%%%%%%%%%%%%%%%%%%%%%%%%%
%%%%%%%%%%%%%%%%%%%%%%%%%%%%%%%%%%%%%%%%%%%

\section{Transmutation Operators}
\label{transm}

Very recently,  a set of first-order differential operator has been proposed \cite{Cheung:2017ems}, which transmutes  amplitudes of various massless theories in arbitrary spacetime dimensions into each other. In this section, we collect some details on these operators, which will be useful in the following.

Scattering amplitudes are functions of Lorentz-invariant products of polarization vectors and momenta\footnote{$v_i v_j$, with $v_i$ a momentum or polarization vector, has to be intended as $v_i \cdot v_j$.}:
\begin{equation}
(e_i  e_j, p_i  e_j, p_i  p_j) \,,
\end{equation}
with the transmutation operators acting on these variables. Since they transmute physical amplitudes into physical amplitudes, they should   preserve on-shell kinematics  and gauge invariance. The following  three types of operators were proposed:\\

\noindent
{\bf Trace operators $\mathcal{T}_{ij}$.}  These two-point operators   reduce the spin of particles $i$ and $j$ by one, placing them within a new color trace structure:
\begin{equation}
\mathcal{T}_{ij} = \partial_{e_ie_j}  \,.
\end{equation}
Therefore they  transmute gravitons into photons, gluons into biadjoint scalars, and BI photons into DBI scalars. They are intrinsically gauge invariant and symmetric $\mathcal{T}_{ij} = \mathcal{T}_{ji}$.
Take for example the $n$-point graviton amplitude $A(h_1,h_2,\dots,h_n)$. The action of the trace operator $\mathcal{T}_{ij}$ will transmute gravitons $i$ and $j$ into photons, which are now within the same trace
\begin{equation}
\label{TO.example1}
\mathcal{T}_{ij} \cdot A(h_1,h_2,\dots,h_n) = A(h_1,h_2,\dots,\gamma_i \, \gamma_j,\dots,h_n)
\end{equation}
which is an amplitude of two photons coupled to $n-2$ gravitons. To outline the trace structure, particles within the same trace are not separated by a comma.\\

\noindent
{\bf Insertion operators $ \mathcal{T}_{ijk}$.} These operators reduce the spin of particle $j$ by one and insert it within an already existing trace structure between particles $i$ and $k$:
\begin{equation}
 \mathcal{T}_{ijk} = \partial_{p_ie_j} - \partial_{p_ke_j} \,.
\end{equation}
 They transmute gravitons into gluons, gluons into biadjoint scalars, and BI photons into pions.
They are not intrinsically gauge invariant but become effectively invariant when combined with proper transmutation operators, i.e. if particles $i$ and $k$ are already transmuted. In particular, the combination $\mathcal{T}_{ijk}\mathcal{T}_{ik}$ is gauge invariant, see \cite{Cheung:2017ems} for more details. The insertion operator is antisymmetric in the first and last index $\mathcal{T}_{ijk} = -\mathcal{T}_{kji}$ and satisfies the additional property $\mathcal{T}_{ijk} + \mathcal{T}_{kjl} = \mathcal{T}_{ijl}$.
As an example, let us consider a color-ordered Yang-Mills amplitude of $n$ gluons and apply, first, the trace operator $\mathcal{T}_{ik}$. Particles $i$ and $k$ are transmuted to biadjoint scalars and placed inside a trace structure with respect to their dual color. The resulting amplitude of $n-2$ gluons and two scalars now exhibits two different trace structures
\begin{equation}
\label{TO.example2}
\begin{split}
\mathcal{T}_{ik} \cdot A(g_1g_2\dots g_n) = A(g_1 \dots,\phi_i  \phi_k, \dots g_n) \,,
\end{split}
\end{equation}
where the biadjoint scalars carry now a dual color, in addition to the original one.
In a second step we apply the insertion operator $\mathcal{T}_{ijk}$ to equation \eqref{TO.example2} transmuting particle $j$ to a biadjoint scalar and inserting it between $i$ and $k$ in the dual color trace. The resulting amplitude reads
\begin{equation}
\mathcal{T}_{ijk}\mathcal{T}_{ik} \cdot A(g_1g_2\dots g_n) = A(g_1 \dots,\phi_i  \phi_j \phi_k, \dots g_n) \,.
\end{equation}

\noindent
It is useful to define the following combination of a single trace operator and various insertion operators 
\begin{equation}
\label{Talpha}
\mathcal{T}[\alpha] = \mathcal{T}_{a_1a_m} \cdot \prod_{k=2}^{m-1}\mathcal{T}_{a_{k-1},a_k,a_m}
\end{equation}
where $\alpha$ with elements $a_k$ is an ordered set. \\

\noindent
{\bf Longitudinal operators $\mathcal{L}_i$.} The one-point longitudinal operators reduce the spin of particle $i$ by one and convert it to a longitudinal mode
\begin{equation}
\label{LOp}
\mathcal{L}_i=\sum_j p_i p_j \partial_{p_j e_i} \,.
\end{equation}
They transmute a graviton into a BI photon, a gluon into a pion and a BI photon into a SG scalar. They can also be written as a linear combination of insertion operators
\begin{equation}
\mathcal{L}_i = \sum_{\substack{j=1 \\ j \neq k}}^n p_ip_j \; \mathcal{T}_{jik} \,,
\end{equation}
 for an arbitrary state $k$.
A  longitudinal operator transmuting all particles of an amplitude  will give a vanishing result.  
The following combination  
\begin{equation} 
\label{type-C}
\mathcal{T}\mathcal{L} = \mathcal{T}_{a_1a_2} \cdot \prod_{\substack{k=1\\k \neq a_1,a_2}}^{n} \mathcal{L}_{k}.
\end{equation}
is gauge invariant  and the resulting amplitude will be a permutation invariant expression of $n$ identical particles. In particular, it will not depend on the choice of the particles $a_1$ and $a_2$. Since the definition \eqref{type-C} is independent of the order of the particles, this type of operator will not induce a trace structure on the resulting amplitude. 
For example, applying \eqref{type-C} to a gravity amplitude we obtain an amplitude of $n$ BI photons
\begin{equation}
A_{\mathrm{BI}}(\gamma_1,\gamma_2, \dots, \gamma_n) = \mathcal{T}\mathcal{L} \cdot A_{\mathrm{G}}(h_1, h_2, \dots, h_n).
\end{equation}
which does not exhibit a trace structure.

The operator \eqref{type-C} can be modified by extending  the single trace operator by a number of insertion operators. Let us start by splitting the set of $n$ particles $O_n = \{ 1,\dots,n\}$ into two disjoint subsets $\alpha$ and $\beta$ such that $\alpha \cup \beta = O_n$, with  $\alpha = \{a_1,a_2,\dots,a_m\}$ and $\beta = \{b_1,\dots,b_{n-m}\}$. We can define the following sequence of  operators
\begin{equation}\label{longitudinal11}
\mathcal{T}[\alpha] \prod_{b \in \beta} \mathcal{L}_b \,,
\end{equation}
where $\mathcal{L}_b$ is the one-point longitudinal operator \eqref{LOp} and $\mathcal{T}[\alpha]$ was defined in \eqref{Talpha}.
We furthermore require $\lvert \alpha \rvert \geq 2$ to ensure that at least one trace operator is contained in $\mathcal{T}[\alpha]$ and $\lvert \beta \rvert > 0$.\\

\begin{figure}[h!]
\centering{
\def\svgwidth{0.7\linewidth}{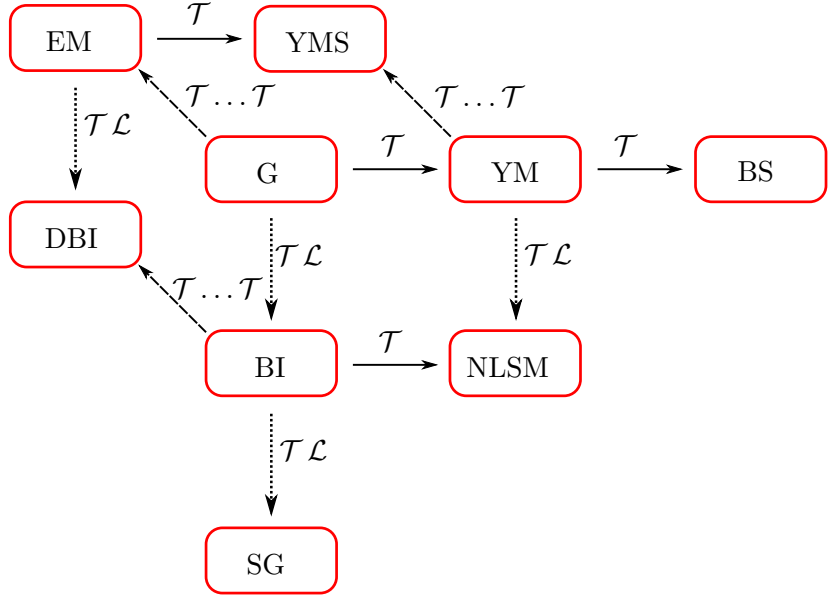}
}
\caption{The connection between different theories through transmutation operators. The theories under consideration are: Gravity (G), Einstein-Maxwell (EM), Yang-Mills (YM), Yang-Mills scalar (YMS), biadjoint scalar (BS), Born-Infeld (BI), Dirac-Born-Infeld (DBI), nonlinear sigma model (NLSM) and special Galileon (SG).}
\label{fig:transm}
\end{figure}

\noindent
In Figure \ref{fig:transm} the action of transmutation operators bringing from one theory to another is shown. 
In particular we have:
 \begin{eqnarray}
 \label{Trasm_ex}
   \left\{
                \begin{array}{ll}
                  A_\mathrm{{BS}} = \mathcal{T}[a_1 \dots a_n] A_{\mathrm{YM}}\\
                A_{\mathrm{YM}} = \mathcal{T}[a_1 \dots a_n] A_{\mathrm{G}}\\
                 A_{\mathrm{NLSM}} = \mathcal{T}[a_1 \dots a_n] A_{\mathrm{BI}}
                \end{array}
              \right.
                 \left\{
                \begin{array}{ll}
                A_{\mathrm{EM}} = \mathcal{T}_{a_1 a_2} \dots \mathcal{T}_{a_{2m-1} a_{2m}} A_{\mathrm{G}}\\
                A_{\mathrm{YMS}} =  \mathcal{T}_{a_1 a_2} \dots \mathcal{T}_{a_{2m-1} a_{2m}} A_\mathrm{{YM}}\\
               A_{\mathrm{DBI}} =  \mathcal{T}_{a_1 a_2} \dots \mathcal{T}_{a_{2m-1} a_{2m}} A_\mathrm{{BI}}
                \end{array}
              \right.
                 \left\{
              \begin{array}{ll}
               A_{\mathrm{SG}} = \mathcal{T} \mathcal{L}  \, A_{\mathrm{BI}} \\
               A_{\mathrm{BI}} =  \mathcal{T} \mathcal{L} \, A_{\mathrm{G}}\\
             A_{\mathrm{NLSM}} = \mathcal{T} \mathcal{L} \, A_{\mathrm{YM}}
                \end{array}
              \right.
                  \end{eqnarray}
By applying  combinations of operators acting only on subset of particles, we can find the amplitudes for extended theories. For instance, \eqref{longitudinal11} generates the extended versions of BI, NLSM and SG.

%%%%%%%%%%%%%%%%%%%%%%%%%%%%%%%%%%%%
%%%%%%%%%%%%%%%%%%%%%%%%%%%%%%%%%%

\section{CHY Formulation}
\label{CHY}

The CHY construction is a compact formulation for tree-level scattering amplitudes of various theories in arbitrary spacetime dimension. In this formulation, the  tree-level scattering amplitude of $n$ massless particles can be expressed as an integral over the moduli space of a $n$-punctured Riemann sphere $\mathfrak{M}_{0,n}$  \cite{Cachazo:2013hca, Cachazo:2014xea}:
\begin{equation}
\label{genericCHY}
A_n^{(0)} =  \int_{\mathfrak{M}_{0,n}}  \frac{\prod_{i=1}^n\mathrm{d}\sigma_i}{\mathrm{vol}\,\mathrm{SL}\left(2,\mathbb{C}\right)} \,\prod_{a}\,' \delta \left( \sum_{\substack{b=1\\ b \neq a}}^n \frac{p_a \cdot p_b}{\sigma_{ab}} \right) \mathcal{I}_n(\{p, e, \tilde{e}, \sigma\})  =:  \int_{\mathfrak{M}_{0,n}} d\mu_{0,n} \, \mathcal{I}_n   \,,
\end{equation}
where $\sigma_i$ denotes the holomorphic coordinate of the puncture $i$ on the Riemann sphere and $\sigma_{ab} = \sigma_a - \sigma_b$.

The integration measure d$\mu_{0,n}$ is a universal object, identical for all theories, and localizes the integral on the so-called scattering equations:
\begin{equation}
\sum_{\substack{b=1\\ b \neq a}}^n \frac{p_a \cdot p_b}{\sigma_{ab}} = 0 \hspace{1cm} \text{for } a\in \{ 1,\dots,n \} \,,
\end{equation}
which relate the kinematic invariants $s_{ab}= 2 p_a \cdot p_b$ and the puncture locations $\sigma_a$.
Originally introduced in  different contexts, i.e. dual resonance models \cite{Fairlie:2008dg} and high-energy behavior of string theory \cite{Gross:1987ar}, they represent an essential object in the formulation \eqref{genericCHY}.  
The prime in the formula indicates that only $n-3$  scattering equations are linearly independent
\begin{equation}
\prod_{a}\,' \delta \left( \sum_{\substack{b=1\\ b \neq a}}^n \frac{p_a \cdot p_b}{\sigma_{ab}} \right) \equiv \sigma_{ij} \sigma_{jk} \sigma_{ki} \prod_{a\neq i,j,k}\, \delta \left( \sum_{\substack{b=1\\ b \neq a}}^n \frac{p_a \cdot p_b}{\sigma_{ab}} \right) \,,
\end{equation}
and, since $\mathrm{dim}({\mathfrak{M}_{0,n}} ) = n-3$, the integral \eqref{genericCHY} is completely localised on $(n-3)!$ of their solutions.

While the form of d$\mu_{0,n}$ is universal, the integrand $\mathcal{I}_n$ depends on the theory under consideration and is a permutation invariant function of the external data $p_a$ and $e_a$, and of the $\sigma$'s.  It exhibits a double-copy structure 
\begin{equation}
\mathcal{I}_n = \mathcal{I}_{L} \mathcal{I}_R
\end{equation}
and, in general, only two building blocks enter its definition. The first one is the Parke-Taylor factor
\begin{equation}
\label{PT}
C_n[\alpha] = \frac{1}{\sigma_{\alpha(1)\alpha(2)}\sigma_{\alpha(2)\alpha(3)}\cdots\sigma_{\alpha(n)\alpha(1)}} \,,
\end{equation}
which depends on the ordering $\alpha$ of the particles contributing to the partial amplitude. 
It appears naturally for every theory exhibiting a color or flavor structure.\\
The second building block is the reduced Pfaffian of an antisymmetric $2n{\times }2n$ matrix $\Psi$ which depends on the external momenta $\{p_1,...p_n\}$ and polarizations $\{e_1,...,e_n\}$ of the particles and on $\{\sigma_1,...,\sigma_n\}$. The matrix $\Psi$ has the following block structure 
\begin{equation}
\label{PsiMatrix}
\Psi = \begin{pmatrix}
A & C  \\
-C^T & B\\
\end{pmatrix}
\end{equation}
where the $n{\times}n$ block matrices are given by 
\begin{equation}
\label{BlockPsi}
A_{ab}=\begin{cases}
\frac{p_a\cdot p_b}{\sigma_{ab}} & a\neq b\\
0 & a=b\\
\end{cases} \hspace{1.5cm}
B_{ab}=\begin{cases}
\frac{e_a\cdot e_b}{\sigma_{ab}} & a\neq b\\
0 & a=b\\
\end{cases}  \hspace{1.5cm}
C_{ab}=\begin{cases}
\frac{p_a\cdot e_b}{\sigma_{ab}} & a\neq b\\
\sum_{c\neq a} \frac{e_a \cdot p_c}{\sigma_{ac}} & a=b\\
\end{cases}
\end{equation}
with $a,b = 1,...,n$. Let us also introduce the following matrix 
\begin{equation}
\label{MatrixX}
\mathcal{X}_{ab}=\begin{cases}
\frac{\delta^{I_a,I_b}}{\sigma_{ab}} & a\neq b\\
0 & a=b
\end{cases} 
\end{equation}
which will appear in the following. $I_a$ denotes the  $U(1)$ charge of particle $a$.

We notice that the Pfaffian of the matrix $\Psi$ is actually vanishing, since the rows and columns are linearly dependent on the solutions of the scattering equations. The reduced Pfaffian $\mathrm{Pf}'\Psi$ is defined via
\begin{equation}
\label{reducedPf}
\mathrm{Pf}'\Psi = \frac{(-1)^{p+q}}{\sigma_{pq}} \mathrm{Pf}(\Psi^{pq}_{pq})
\end{equation}
for some $1\le p < q \le n$, and where $\Psi^{pq}_{pq}$ is the matrix $\Psi$ with the rows and columns $p$ and $q$ removed.

For instance, the CHY integrand for an $n$-point gravity amplitude is given by 
\begin{equation}
\label{CHYgr}
\mathcal{I}_\mathrm{G}(p,e,\tilde{e},\sigma) = \mathrm{Pf'}\Psi(p,e,\sigma) \; \mathrm{Pf'}\Psi(p,\tilde{e},\sigma)
\end{equation}
where $e_a$ and $\tilde{e}_a$ are two sets of polarization vectors which, together,  describe  the polarization tensor of the state $a$: $\epsilon_a^{\mu\nu} = e_a^\mu \tilde{e}^\nu_a$. This gravity theory, also called NS-NS gravity, describes gravitons, dilatons and B-field states and it is the theory descending from YM via KLT relations. To retrieve pure Einstein-gravity amplitudes, an appropriate symmetric, traceless linear combination of $e_a^\mu \tilde{e}^\nu_a$ terms should be chosen.
The integrand for the Yang-Mills tree-level partial amplitude with ordering  $[\alpha]$ is instead
\begin{equation}
\mathcal{I}_{\mathrm{YM}, \alpha}(p,e,\sigma) = C_n[\alpha] \; \mathrm{Pf'}\Psi(p,\tilde{e},\sigma) \,,
\end{equation}
and for the full amplitude we can write
\begin{equation}
C_n = \sum_{\alpha \in S_n/Z_n} \, \mathrm{Tr}\left(T^{a_{\alpha{(1)}}} \, T^{a_{\alpha{(2)}}} \dots T^{a_{\alpha{(n)}}}\right)  C_n[\alpha] \,.
\end{equation}
In Table \ref{table:integrands} we summarize the various integrands which will appear in this paper  \cite{Cachazo:2013iea, Cachazo:2014xea}.

\bgroup
\def\arraystretch{1.7}
\begin{table}[h!]
\begin{center}
  \begin{tabular}{ c | c | c}
    \hline
    {\bf Theory} & {\bf   $\mathcal{I}_L$}  & {\bf   $\mathcal{I}_R$} \\ \hline
    Gravity (G) &  $ \mathrm{Pf'\Psi}(p,e,\sigma) $ & $ \mathrm{Pf'\Psi}(p,\tilde e,\sigma)$ \\ \hline
    Yang-Mills (YM) &  $C_n[\alpha]  $ & $ \mathrm{Pf'\Psi}(p,\tilde e,\sigma)$ \\ \hline
    Bi-adjoint Scalar (BS) &  $C_n[\alpha] $ & $ C_n[\beta]$ \\   \hline
    Einstein-Maxwell  (EM) &  $\mathrm{Pf} \mathcal{X}(\sigma) \, \mathrm{Pf}'A(p,\sigma) $ & $ \mathrm{Pf'}{\Psi}(p,\tilde e,\sigma)$ \\   \hline
    Born-Infeld (BI) &  $[\mathrm{Pf}'A(p,\sigma)]^2 $ & $ \mathrm{Pf'\Psi}(p,\tilde e,\sigma)$ \\   \hline
    Dirac-Born-Infeld (DBI) &  $[\mathrm{Pf}'A(p,\sigma)]^2 $ & $\mathrm{Pf}\mathcal{X}(\sigma) \, \mathrm{Pf}'A(p,\sigma) $ \\   \hline
    Yang-Mills-scalar (YMS) &  $C_n[\alpha] $ & $\mathrm{Pf} \mathcal{X}(\sigma) \,\mathrm{Pf}'A(p,\sigma) $ \\   \hline
    Non-linear sigma model (NLSM) &  $C_n[\alpha] $ & $[\mathrm{Pf}'A(p,\sigma)]^2 $ \\   \hline
    Special Galileon (SG) &  $[\mathrm{Pf}'A(p,\sigma)]^2 $ & $ [\mathrm{Pf}'A(p,\sigma)]^2$ \\   \hline
      \end{tabular}
       \caption{CHY integrands for various theories. The definitions of $C_n[\alpha]$, $\Psi$, $A$ and $\mathcal{X}$ can be found in eq.s \eqref{PT}, \eqref{PsiMatrix}, \eqref{BlockPsi}, \eqref{MatrixX} respectively. }
    \label{table:integrands}
    \end{center}
  \end{table}
  \egroup

%%%%%%%%%%%%%%%%%%%%%%%%%%%%%%%%%%%%
%%%%%%%%%%%%%%%%%%%%%%%%%%%%%%%%%%

\section{Transmutation Operators acting on CHY Integrands}
\label{transm_CHY}
In this section we are going to apply the transmutation operators to the CHY integrands and retrieve all theories which have been described in the previous part of the paper, plus their extensions. Since the following paragraphs are technical in nature, we encourage the reader to look at the figure at the end of the section, where we collect all our results.

\subsection{Single Trace Amplitudes: G $\rightarrow$ YM $\rightarrow$ BS and BI $\rightarrow$ NLSM}

Let us start by studying the action of an operator $\mathcal{T}[\alpha]$ as in \eqref{Talpha} on the reduced Pfaffian $\mathrm{Pf'}\Psi$ for an arbitrary number of particles $n$. Let $\alpha$ be an ordered subset of $\{1,\dots,n\}$ with $m \leq n$ elements. The operator consists of a single trace operator and $m-2$ insertion operators:
\begin{equation}
\mathcal{T}[\alpha] = \mathcal{T}_{a_1a_m} \cdot \prod_{k=2}^{m-1}\mathcal{T}_{a_{k-1},a_k,a_m} \,.
\end{equation}

First, we consider the case $m=n$, where all particles are transmuted. Without loss of generality, we consider the canonical ordering $1,2,\dots,n$ and choose to delete rows and columns $p=1$ and $q=2$ in the definition of the reduced Pfaffian \eqref{reducedPf}. This is not a restriction, as the result can be easily generalized to different orderings and the definition of the reduced Pfaffian is independent of $p$ and $q$. The object that needs to be computed then is
\begin{equation}\label{st_easyop}
\mathcal{T}[12\dots n] \cdot \mathrm{Pf}'\Psi = \mathcal{T}_{1n} \cdot \prod_{a=2}^{n-1}\mathcal{T}_{a-1,a,n} \cdot \mathrm{Pf}'\Psi  \,.
\end{equation}

To determine the action of the transmutation operators let us rewrite them in terms of the components $\Psi_{ab}$, while keeping in mind that the indices 1 and 2 do not appear in the reduced Pfaffian. We will consider first the trace operator $\mathcal{T}_{1n}$ and afterwards the insertion operators $\mathcal{T}_{ijk}$. In the latter, $\mathcal{T}_{12n}$ and $\mathcal{T}_{23n}$ have to be treated separately due to the choice we made of $p=1$ and $q=2$.
\begin{itemize}
\item $\mathcal{T}_{1n}=\partial_{e_1 e_n}$: This operator will act on the component $\Psi_{n+1,2n} = B_{1n} = \frac{e_1e_n}{\sigma_{1n}}$ of \eqref{PsiMatrix}, see \eqref{BlockPsi}.  The action on  $B_{n1}$ is irrelevant for the purpose of computing the Pfaffian, since only one term will appear. Therefore, we can rewrite 
\begin{equation}
\mathcal{T}_{1n} = \frac{1}{\sigma_{1n}} \frac{\partial}{\partial\Psi_{n+1,2n}} \,.
\end{equation}
%%%%%%%
\item $\mathcal{T}_{a-1,a,n}  = \partial_{p_{a-1} e_a} - \partial_{p_n e_a}$, $a \in \{4,...,n-1\}$ : The variable $p_{a-1}e_a$ appears in the two components $\Psi_{a-1,a+n} = C_{a-1,a} = \frac{p_{a-1}e_a}{\sigma_{a-1,a}}$ and $\Psi_{a,a+n} = C_{aa} = \sum_{b\neq a} \frac{p_be_a}{\sigma_{ab}}$. The latter component contains also $p_{n}e_a$, which further appears in $\Psi_{n,a+n} = C_{na} = \frac{p_ne_a}{\sigma_{na}}$. In combination this leads to
$$\mathcal{T}_{a-1,a,n} = \frac{1}{\sigma_{a-1,a}} \frac{\partial}{\partial\Psi_{a-1,a+n}} + \frac{1}{\sigma_{an}} \frac{\partial}{\partial\Psi_{n,a+n}} - \left( \frac{1}{\sigma_{a-1,a}} + \frac{1}{\sigma_{an}} \right) \frac{\partial}  {\partial\Psi_{a,a+n}}$$
This formula can be simplified when $a=2,3$ as the components with indices 1 and 2 do not appear in the reduced Pfaffian: 
%%%%%%%%%
\item $\mathcal{T}_{23n}= \partial_{p_2 e_3} - \partial_{p_n e_3}$: 
  This corresponds to the case $a=3$. We can simply ignore the first term, which would have a derivative with respect to $\Psi_{2,n+3}$, to find
$$\mathcal{T}_{23n} = \frac{1}{\sigma_{3n}} \frac{\partial}{\partial\Psi_{n,n+3}} - \left( \frac{1}{\sigma_{23}} + \frac{1}{\sigma_{3n}} \right) \frac{\partial}  {\partial\Psi_{3,n+3}}.$$
%%%%%%%%
\item $\mathcal{T}_{12n}= \partial_{p_1 e_2} - \partial_{p_n e_2}$ : In the case $a=2$, we can ignore the first ($\Psi_{1,n+2}$) and the last term ($\Psi_{2,n+2}$), 
$$\mathcal{T}_{12n} = \frac{1}{\sigma_{2n}} \frac{\partial}{\partial\Psi_{n,n+2}}.$$
\end{itemize}
We note that every transmutation operator can be represented as a linear differential operator in the components $\Psi_{ab}$, replacing a certain component with a corresponding $\sigma_{ab}$. 
We are now ready to act on the reduced Pfaffian.
The reduced Pfaffian consists of $(2n-3)!!$ terms,  with each of them consisting of $n-1$ factors, see Appendix \ref{app:pfaffian}. Crucially, in every term each of the $2n-2$ indices appears exactly one time. Furthermore, every term has a different combination of indices. By acting with a transmutation operator on the reduced Pfaffian, only those terms containing the exact same index combination will survive. As we discussed above, each component will be replaced by the corresponding combination of $\sigma$'s. Repeating this for all $n-1$ operators, the number of relevant terms will decrease and eventually all components will be replaced. Let us show how this works in detail.

Since the transmutation operators commute among themselves, we can choose their ordering. We start with the simplest one, namely $\mathcal{T}_{1n}$, which makes most of the terms vanish and keeps only those that involve the component $\Psi_{n+1,2n}$. This component is then replaced by the factor $\frac{1}{\sigma_{1n}}.$ 
The second operator we consider is $\mathcal{T}_{12n}$. It limits the number of relevant terms further, to those containing also the component $\Psi_{n,n+2}$. The next operator in the process shall be $\mathcal{T}_{23n}$. Its first term involves a derivative with respect to $\Psi_{n,n+3}$ but, since every index appears only once per term, there are no terms with index $n$ left after having acted with $\mathcal{T}_{12n}$. Therefore only the derivative w.r.t $\Psi_{3,n+3}$ is relevant. 
Finally, we repeat this procedure with the remaining three-point operators $\mathcal{T}_{a-1,a,n}$ for $a \in \{4,\dots,n-1\}$. Even though they contain the three pairs of indices $(a-1,a+n)$, $(n,a+n)$ and $(a,a+n)$,  for every $a$ only the last pair is relevant. Indeed the second term again involves the index $n$ which was already deleted by $\mathcal{T}_{12n}$, while the first term contains the index $a-1$ which was deleted by the third term of $\mathcal{T}_{a-2,a-1,n}$.

Therefore, we have showed that there is only a single term in the Pfaffian on which the $n-1$ transmutation operators act non-trivially. 
The action of the complete transmutation operator on the Pfaffian is given by
\begin{equation}
\begin{split}
\mathcal{T}[12...n] \cdot \mathrm{Pf}'(\Psi) = & \, s_n \frac{1}{\sigma_{12}} \frac{1}{\sigma_{n1}}\frac{1}{\sigma_{2n}} \left( \frac{1}{\sigma_{23}} + \frac{1}{\sigma_{3n}} \right) \prod_{a=4}^{n-1} \left( \frac{1}{\sigma_{a-1,a}} + \frac{1}{\sigma_{an}} \right) \\[1em]
= & \, s_n \, \frac{1}{\sigma_{12}\cdots\sigma_{n1}}
\end{split}
\end{equation}
where the factor $\frac{1}{\sigma_{12}}$ comes from the fact that we have chosen to remove rows and columns $\{1,2\}$, see \eqref{reducedPf}.  
$s_n$ is a sign factor which depends on $n$ 
\begin{equation}
s_n=  (-1)^{\frac{n(n+1)}{2}-1}
\end{equation}
and is calculated by considering the sign coming from the Pfaffian, see Appendix \ref{app:pfaffian}, from the operators and the reduced Pfaffian \eqref{reducedPf}.
We see that acting with the transmutation operator on the reduced Pfaffian, we are left with the Parke-Taylor factor corresponding to the chosen ordering:
\begin{equation}
\mathcal{T}[\alpha] \,
 \mathrm{Pf'} \Psi =  s_n C_n[\alpha] \,.
\end{equation}

We are now ready to transmute scattering amplitudes.
Acting with the operator \eqref{st_easyop} on the CHY integrand of gravity \eqref{CHYgr} we obtain the integrand for YM theory ordered with respect to the set $\alpha$. Specifically, up to an irrelevant sign, 
\begin{equation}
\label{YMintegrand}
 \mathcal{T}[\alpha] \, \mathcal{I}_{\mathrm{G}} =  \mathcal{T}[\alpha] \,\left( \mathrm{Pf'}\tilde{\Psi} \mathrm{Pf'} \Psi  \right)= 
\frac{1}{\sigma_{\alpha(1)\alpha(2)}\cdots\sigma_{\alpha(n)\alpha(1)}}\;\mathrm{Pf'}\Psi = \mathcal{I}_{\mathrm{YM}}(\alpha) \,.
\end{equation}
Acting again with the operator in an ordering $\beta$, we obtain a canonically-ordered integrand for BS theory in the dual ordering with respect to $\alpha$ and $\beta$
\begin{equation}
\label{BSintegrand}
 \mathcal{T}[\beta]\, \mathcal{I}_{\mathrm{YM}}(\alpha)  = \mathcal{T}[\beta] \, \left( C_n[\alpha] \mathrm{Pf'}\Psi \right) =  C_n[\alpha] \, \frac{1}{\sigma_{\beta(1)\beta(2)}\cdots\sigma_{\beta(n)\beta(1)}} =\mathcal{I}_{\mathrm{BS}}(\alpha|\beta) \,.
\end{equation}
In the same way, a BI amplitude is transmuted into a NLSM one:
\begin{equation}
\label{BIintegrand}
 \mathcal{T}[\alpha]\, \mathcal{I}_{\mathrm{BI}} =  \mathcal{T}[\alpha]  \left(\mathrm{Pf'} \Psi(\mathrm{Pf' A})^2\right) = \frac{1}{\sigma_{\alpha(1)\alpha(2)}\cdots\sigma_{\alpha(n)\alpha(1)}} \, (\mathrm{Pf' A})^2 =\mathcal{I}_{\mathrm{NLSM}}(\alpha) \,.
\end{equation}

These integrands agree with \cite{Cachazo:2013iea}.
 It is straightforward to extend the results above to the cases when not every  particle is transmuted, i.e. $m<n$, and find the integrands for extended theories. 
 \begin{equation}
 \label{gen_st}
\mathcal{T}[\alpha] \,
 \mathrm{Pf'} \Psi =  s_{m,n} C_m[\alpha] \, {\mathrm{Pf}}[\Psi]_{\bar\alpha;\bar\alpha} \,,
\end{equation}
where the sign $s_{m,n} = (-1)^{\frac{m(m+1)}{2}-1+n(m+1)}$  now depends also on $m$ and $\bar\alpha$ is the $n-m$ subset of non-transmuted particles. $[\Psi]_{\bar{\alpha};\bar{\alpha}}$ is therefore the $2(n-m){\times}2(n-m)$ matrix that is obtained by restricting $\Psi$ to the subset $\bar{\alpha}$.

%%%%%%%%%%%%%%%%%%%%%%%%%%%%%%%%%%%%%%%%%%%%%%%%%%%%%%%%%%%%%%%%
%%%%%%%%%%%%%%%%%%%%%%%%%%%%%%%%%%%%%%%%%%%%%%%%%%%%%%%%%%%%%%%%

\subsection{Multiple Trace Amplitudes: G $\rightarrow$ EM and YM $\rightarrow$ YMS and BI $\rightarrow$ DBI}

Let us study now the action of  transmutation operators on CHY integrands  for multiple trace  amplitudes. In particular, we start by considering a simple sequence of $m$ trace operators 
\begin{equation}
\label{MTO}
\mathcal{T}_{\{\alpha\beta\}}^m = \prod_{k=1}^m \mathcal{T}_{a_kb_k}  \,,
\end{equation}
transmuting pairwise a subset of $2m$ particles  
$\{ \alpha\beta \} = \{ a_1,b_1; a_2,b_2; \dots; a_m,b_m\}$ with $a_k < b_k$\footnote{This can be done w.l.o.g. due to the symmetry properties of $\mathcal{T}_{ab}$}. We consider $2m \leq n$, therefore not all particles are necessarily transmuted. Due to permutation invariance, we can define the reduced Pfaffian $\mathrm{Pf}' \Psi$ such that all particles from the set $\{ \alpha\beta \}$ are placed in the last $2m$ rows and columns. As a starting point, we consider the action of a single trace operator $\mathcal{T}_{ab}$ with $a<b$ on the reduced Pfaffian $\mathrm{Pf}'(\Psi)$.
The variable $e_a e_b$ appears only once and corresponds to the matrix entry $\Psi_{a+n,b+n} = B_{ab} = \frac{e_ae_b}{\sigma_{ab}}$. Using the recursive definition of the Pfaffian \eqref{recursivePfaffian} with  $i=a+n$, we find
\begin{equation}
\mathrm{Pf}'(\Psi) = \frac{(-1)^{p+q}}{\sigma_{pq}} \sum_{\substack{j=1\\j\neq p,q\\j\neq a+n}}^{2n}(-1)^{n+a+j+1+\Theta(n+a-j)} \, \Psi_{n+a,j} \, \mathrm{Pf}(\Psi_{p,q,n+a,j}^{p,q,n+a,j}) \,.
\end{equation}
The product $e_a e_b$ does not appear in the remaining Pfaffian, because the row and column $n+a$ were erased, while  it  is present once in $\Psi_{n+a,j}$ for $j=n+b$. Therefore we can write 
\begin{equation}
\label{EM2}
\mathrm{Pf}'(\Psi) = (-1)^{a+b+1+\Theta(a-b)} \, \frac{e_a e_b}{\sigma_{ab}} \, \mathrm{Pf}'(\Psi_{n+a,n+b}^{n+a,n+b}) + ... \,,
\end{equation}
where the ellipses indicate terms vanishing after the action of  the transmutation operator.
Finally we find 
\begin{equation}
\label{EM1}
\mathcal{T}_{ab} \cdot \mathrm{Pf}' (\Psi) = \frac{(-1)^{a+b+1}}{\sigma_{ab}} \mathrm{Pf}'(\Psi_{a+n,b+n}^{a+n,b+n}) \,.
\end{equation}
Let us proceed by studying the case $m=2$, where another trace operator $\mathcal{T}_{cd}$ with $c<d$ and $a$, $b$, $c$, $d$ pairwise different is applied to equation \eqref{EM1}. Using again the recursion relation for $i=c+n$ we can write
\begin{equation}
\mathrm{Pf}'(\Psi_{a+n,b+n}^{a+n,b+n}) = (-1)^{c+d+1} \, \mathrm{sgn}(abcd) \, \frac{e_c e_d}{\sigma_{cd}} \, \mathrm{Pf}' (\Psi_{n+a,n+b,n+c,n+d}^{n+a,n+b,n+c,n+d}) + ... \,,
\end{equation}
where again all terms denoted by the dots vanish when applying $\mathcal{T}_{cd}$. Here $\mathrm{sgn}(abcd)$ is the sign of the permutation  $\{a,b,c,d\}$. Therefore, one can write for a sequence of two trace operators acting on the reduced Pfaffian
\begin{equation}
\label{EM3}
\mathcal{T}_{cd}\mathcal{T}_{ab} \cdot \mathrm{Pf}' (\Psi) = \frac{(-1)^{a+b+c+d} \, \mathrm{sgn}(abcd)}{\sigma_{ab}\sigma_{cd}} \mathrm{Pf}'(\Psi_{a+n,b+n,c+n,d+n}^{a+n,b+n,c+n,d+n}).
\end{equation}
We are now ready to generalize the result to the full operator \eqref{MTO}, by using \eqref{EM1} and \eqref{EM3}: 
\begin{equation}
\label{EM4}
\mathcal{T}_{\{\alpha\beta\}}^m\mathrm{Pf}' (\Psi) =  \frac{\mathrm{sgn}(\{ \alpha\beta \} )}{\prod_{k=1}^m \sigma_{a_k,b_k}} \, 
 \mathrm{Pf}'\left( \Psi_{;\{ \alpha\beta \}}^{;\{ \alpha\beta \}} \right) \,,
\end{equation}
where the notation $\Psi_{;\{ \alpha\beta \}}^{;\{ \alpha\beta \}}$ indicates that we removed all rows and columns corresponding to $\{ \alpha\beta \}$ from the second block of $\Psi$ and left the first block unchanged. The factor $\mathrm{sgn}(\{ \alpha\beta \} )$ is the sign of the permutation
\begin{equation}
\sigma = 
\begin{pmatrix}
a_1 & b_1 & a_2 & b_2 & \dots & a_m & b_m\\
\hat{a}_1 & \hat{b}_1 & \hat{a}_2 & \hat{b}_2 & \dots & \hat{a}_m & \hat{b}_m
\end{pmatrix}
\end{equation}
with $\hat{a}_1 < \hat{b}_1 < \hat{a}_2 < \hat{b}_2 < \dots < \hat{a}_{m} < \hat{b}_m$ . 
Let us note that \eqref{EM4} can be rewritten in terms of  the matrix $\mathcal{X}$ \eqref{MatrixX}. 
With $(\mathcal{X})_{\{\alpha\beta\}}$ we denote the reduced $2m \times 2m$ matrix that is obtained from
$\mathcal{X}$ by deleting all rows and columns that correspond to non-transmuted particles.
Recalling that the operator \eqref{MTO} transmutes states pairwise regarding the
color structure, we find for the color charges of the transmuted particles $I_{a_{j}} = I_{b_{k}}$ if and only if $j=k$. In this special case, 
\begin{equation}
\frac{\mathrm{sgn}(\{ \alpha\beta \} )}{\prod_{k=1}^m \sigma_{a_k,b_k}} = \mathrm{Pf}(\mathcal{X})_{\{\alpha\beta\}} \,,
\end{equation}
and therefore we can rewrite \eqref{EM4} 
\begin{equation}
\mathcal{T}_{\{\alpha\beta\}}^m \mathrm{Pf}' (\Psi) =  \mathrm{Pf}(\mathcal{X})_{\{\alpha\beta\}} \,  
 \mathrm{Pf}'\left( \Psi_{;\{ \alpha\beta \}}^{;\{ \alpha\beta \}} \right) \,.
\end{equation}

\noindent
To get a better understanding of equation \eqref{EM4}, let us investigate the special case where the highest possible number of trace operators is applied to the reduced Pfaffian. Since every trace operator carries two indices, at most $\lfloor \frac{n}{2} \rfloor$ operators can be applied. There are two different cases:
\begin{itemize}
\item $n$ even: Exactly $m = \frac{n}{2}$ trace operators can be applied. Every product of the form $e_ae_b$ is erased from the matrix $\Psi$ and only the upper left block $A$ survives inside the Pfaffian, see \eqref{PsiMatrix}. The result does not depend anymore on the external polarizations and 
\eqref{EM4} reduces to
\begin{equation}\label{EM_even}
\mathcal{T}_{\{\alpha\beta\}}^m\mathrm{Pf}' (\Psi) = \frac{\mathrm{sgn}(\{ \alpha\beta \} )}{\prod_{k=1}^{\frac{n}{2}} \sigma_{a_k,b_k}} \mathrm{Pf}'(A).
\end{equation}

\item $n$ odd: The maximal number of trace operators which can be applied is $m = \frac{n-1}{2}$, with one of the $n$ labels  not appearing in $\{ \alpha\beta \}$. Let us denote this label by $f$. In the lower right block of $\Psi$, the submatrix $B$, all rows and columns are erased apart from one. The single entry remaining is $B_{ff} = 0$. In the off-diagonal blocks, a single row or respectively column survives. 
We then obtain
\begin{equation}
\label{EM_odd1}
\mathcal{T}_{\{\alpha\beta\}}^m \, \mathrm{Pf}' (\Psi) = \frac{(-1)^{f+1} \, \mathrm{sgn}(\{ \alpha\beta \} )}{\prod_{k=1}^{\frac{n-1}{2}} \sigma_{a_k,b_k}} \mathrm{Pf}'\begin{pmatrix} A & C_{x,f} \\ (-C_{x,f})^T & 0 \end{pmatrix}
\end{equation}
where $C_{x,f}$ denotes the $f$-th column of $C$. By using the recursive definition of the Pfaffian \eqref{recursivePfaffian} for $i=n+1$ this can be further rewritten as
\begin{equation}
\label{EM_odd2}
\mathcal{T}_{\{\alpha\beta\}}^m \cdot \mathrm{Pf}' (\Psi) = \frac{(-1)^{f+1} \, \mathrm{sgn}(\{ \alpha\beta \} )}{\prod_{k=1}^{\frac{n-1}{2}} \sigma_{a_k,b_k}} \; \sum_{\substack{j=1\\j \neq p,q}}^n (-1)^{j+1} C_{j,f} \mathrm{Pf}'(A_j^j).
\end{equation}
where $p$ and $q$ denote the two rows and columns that are erased when taking the reduced Pfaffian.
\end{itemize}
We are now ready to evaluate the CHY integrands for Einstein-Maxwell theory and Yang-Mills scalar theory. Let us start from gravity \eqref{CHYgr}
and apply the chain of operators \eqref{MTO} 
\begin{eqnarray}\label{EM5}
\mathcal{I}_{\mathrm{EM}}(\gamma_{a_1}\gamma_{b_1},\dots,\gamma_{a_m}\gamma_{b_m};\{ h\} ) &=&  \mathcal{T}_{\{\alpha\beta\}}^m \mathcal{I}_{\mathrm{G}} \nonumber \\
&=& \mathrm{Pf}(\mathcal{X})_{\{\alpha\beta\}} \, \mathrm{Pf}'\left( \Psi_{;\{ \alpha\beta \} }^{;\{ \alpha\beta \} }(p,e,\sigma) \right) \, \mathrm{Pf}'\Psi(p,\tilde{e},\sigma).
\end{eqnarray}
Equation \eqref{EM5} is the expression for the CHY integrand of an EM theory with $2m$ photons $\gamma_a$ and $n-2m$ gravitons. Each trace operator applied above transmutes two gravitons into two photons and puts them into a separate color trace. Hence the integrand contains $m$ traces of two photons,  separated in equation \eqref{EM5} by a comma, and  the remaining $\{ h \}$ gravitons.
This result for the EM integrand agrees with the results from \cite{Cachazo:2014xea} when appropriately modified to have photons arranged in $m$ separate pairs of color traces.
The $U(1)$ charges $I$ are chosen such that $I_{a_k} = I_{b_k} $ for all $ k$ and different otherwise.
In the case $n=2m$, where all particles are transmuted, the matrix inside the Pfaffian reduces to the matrix $A$, see \eqref{EM_even}.   
The CHY integrand for YMS can be derived from the Yang-Mills integrand  \eqref{YMintegrand}

\begin{eqnarray}\label{YMS}
\mathcal{I}_{\mathrm{YMS}}^{[\alpha]}(\phi_{a_1}\phi_{b_1},\dots,\phi_{a_m}\phi_{b_m};\{ g\} ) &=& \mathcal{T}_{\{\alpha\beta\}}^m\cdot \mathcal{I}_{\mathrm{YM}}^{[\alpha]}  \nonumber \\
&=&\mathrm{Pf}(\mathcal{X})_{\{\alpha\beta\}} \, \mathrm{Pf}'\left( \Psi_{;\{ \alpha\beta \} }^{;\{ \alpha\beta \} }(p,e,\sigma) \right) \, C_n[\alpha].
\end{eqnarray}
Equation \eqref{YMS} is the expression for the CHY integrand of a YMS theory with $2m$ scalars and $n-2m$ gluons  $\{ g \}$. Similarly to the previous case, since every trace operator transmutes two gluons into two scalars in a separate trace, the integrand consists of $m$ traces containing two scalars each, in agreement with \cite{Cachazo:2014xea}.
Finally, the integrand for DBI can be obtained from the BI theory, which we will derive in the next section
\begin{eqnarray}\label{DBI}
\mathcal{I}_{\mathrm{DBI}}(\phi_{a_1}\phi_{b_1},\dots,\phi_{a_m}\phi_{b_m};\{ \gamma \} ) &=& \mathcal{T}_{\{\alpha\beta\}}^m \cdot \mathcal{I}_{\mathrm{BI}}(\gamma_1,\dots,\gamma_n) \nonumber\\
&=& \mathrm{Pf}(\mathcal{X})_{\{\alpha\beta\}} \, \mathrm{Pf}'\left( \Psi_{;\{ \alpha\beta \} }^{;\{ \alpha\beta \} }(p,e,\sigma) \right) \, \left[ \mathrm{Pf}'A(p,\sigma) \right]^2.
\end{eqnarray}

Equation \eqref{DBI} describes the CHY integrand of a theory with $2m$ DBI scalars coupled to $n-2m$ BI photons $\{ \gamma \}$ and agrees with \cite{Cachazo:2014xea} when a similar discussion as for the EM charges is done. The scalars are organized in $m$ pairs, each one in a different trace.

%%%%%%%%%%%%%%%%%%%%%%%%%%%%%%%%%%%%%%%%%%%%%%%%%%%%%%%%%%%%%%%%
%%%%%%%%%%%%%%%%%%%%%%%%%%%%%%%%%%%%%%%%%%%%%%%%%%%%%%%%%%%%%%%%

\subsection{Longitudinal Operators: G $\rightarrow$ BI $\rightarrow$ SG and YM $\rightarrow$ NLSM}

Finally, we would like to calculate the  CHY integrands for BI, NLSM and SG theories by using the longitudinal operators. Let us remind that we can define the following sequence of  operators
\begin{equation}
\label{longitudinal1}
\mathcal{T}[\alpha] \prod_{b \in \beta} \mathcal{L}_b \,,
\end{equation}
where $\mathcal{L}_b$ is the one-point longitudinal operator \eqref{LOp} and $\alpha \cup \beta$ covers all the set of $n$ particles of $O_n$.
As explained in \cite{Cheung:2017ems}, we can rewrite a product of longitudinal operators in the following way, for $\lvert \beta \rvert$ even:
\begin{equation}
\label{expansion}
\prod_{b \in \beta} \mathcal{L}_{b} = \sum_{\rho \in P_{n-m}} \prod_{k=1}^{\frac{n-m}{2}} \mathcal{L}_{i_kj_k} + \cdots
\end{equation}
where $\mathcal{L}_{ij}$ is the two-point longitudinal operator 
\begin{equation}
\mathcal{L}_{ij} = - p_ip_j \partial_{e_ie_j}
\end{equation}
and the ellipses denote remainder terms which vanish when transmuting all particles of a physical amplitude. The sum is over the set of partitions of $\beta$ into pairs.
For $\lvert \beta \rvert$ odd, we can exclude one arbitrary element -- call it $x$ -- from $\beta$ to obtain a new set $\beta'$. Then
\begin{equation}
\label{expansionOdd}
\prod_{b \in \beta} \mathcal{L}_{b} = \mathcal{L}_x \cdot \prod_{b \in \beta'} \mathcal{L}_{b}
\end{equation}
and use \eqref{expansion}  for the reduced set $\beta'$. 
This expansion simplifies the calculations as the two-point operator contains the same derivative as the trace operator we have already discussed.

The number $\lvert \alpha \rvert$ determines the fraction of particles which are not transmuted by the longitudinal operators. The case $\lvert \alpha \rvert = 2$ directly connects  to \cite{Cheung:2017ems}: in this case the operator transmutes amplitudes from gravity to Born-Infeld, from Born-Infeld to Special Galileon and from Yang-Mills to the NLSM. For $\lvert \alpha \rvert > 2$ the operator yields amplitudes from extended theories \cite{Cachazo:2016njl} and it transmutes 
\begin{itemize}
\item Gravity amplitudes into amplitudes of BI photons coupled to gluons
\item BI amplitudes into amplitudes of SG scalars coupled to pions
\item Gluon amplitudes into pions coupled to biadjoint scalars
\end{itemize}

\noindent
Let us start with the case $\lvert \alpha \rvert = 2$, i.e. the operator
\begin{equation}\label{longitudinal_op}
\mathcal{T}_{a_1a_2} \cdot \prod_{k=1}^{n-2} \mathcal{L}_{b_k} \,.
\end{equation}
It consists of a single trace operator, whose action is already known from the previous section, and $n-2$ longitudinal operators. These can be expanded using \eqref{expansion}:

\begin{itemize}
%%%%%%%%%%%%%%%%%%%
\item $n$  even: 
we can use \eqref{expansion} to rewrite the longitudinal operators as a sum of products of $r=\frac{n}{2}-1$ two-point longitudinal operators. The action of one particular partition on the reduced Pfaffian is given by equation \eqref{EM_even} multiplied by the products of momenta coming from the longitudinal operators:
\begin{equation}\label{longitudinal2}
\mathcal{T}_{a_1a_2}\cdot \mathcal{L}_{i_1j_1} \cdots \mathcal{L}_{i_rj_r} \mathrm{Pf}'(\Psi) = \frac{(-1)^r \, \mathrm{sgn}(a_1,a_2,\{ ij \})}{\sigma_{a_1a_2}} \prod_{k=1}^{r} \left( A_{i_kj_k} \right) \mathrm{Pf}'(A)
\end{equation}
where $A_{ij} = \frac{p_ip_j}{\sigma_{ij}}$ is an element of the matrix $A$.
The sign is determined by the ordering of all $i$ and $j$ as well as $a_1$ and $a_2$, which together cover the full set $\{1,\dots,n\}$. However, the positions of $a_1$ and $a_2$ are the same for every partition and hence the sign factorizes into
\begin{equation}
\mathrm{sgn}(a_1,a_2,\{ ij \}) = (-1)^{a_1+a_2+1}\, \mathrm{sgn}(\{ij \}) \,.
\end{equation} 
Taking the sum over all possible partitions of $\beta$ into pairs turns the product in equation \eqref{longitudinal2}, together with the sign, into the Pfaffian of the matrix $A_{a_1,a_2}^{a_1,a_2}$. Recalling the definition of the reduced Pfaffian $\mathrm{Pf}'\Psi = \frac{(-1)^{p+q}}{\sigma_{pq}}\mathrm{Pf}(\Psi_{pq}^{pq})$ this becomes
\begin{equation}
\mathcal{T}_{a_1a_2} \cdot \prod_{k=1}^{n-2} \mathcal{L}_{b_k} \cdot \mathrm{Pf}'(\Psi) = (-1)^\frac{n}{2}\mathrm{Pf}'(A)^2 \,.
\end{equation}
%%%%%%%%%%%%%%%%
\item $n$ odd:
using \eqref{expansionOdd} we rewrite the longitudinal operators as a sum of products of $r=\frac{n-3}{2}$ two-point longitudinal operators multiplied by a single one-point longitudinal operator $\mathcal{L}_x$. In analogy to equation \eqref{EM_odd2}, the action of the $r$ two-point operators together with the trace operator is given by 
\begin{equation}
\label{longitudinal_odd}
\begin{split}
\mathcal{T}_{a_1a_2} \cdot \mathcal{L}_{i_1j_1} \cdots \mathcal{L}_{i_rj_r} \cdot \mathrm{Pf}' (\Psi) =  \frac{(-1)^{r+x+1} \, \mathrm{sgn}(\{ ij \} )}{\sigma_{a_1a_2}}\prod_{k=1}^{r} \left( A_{i_k,j_k} \right) 
\sum_{\substack{j=1\\j \neq p,q}}^n (-1)^{j+1} C_{jx} \mathrm{Pf}'(A_j^j).
\end{split}
\end{equation}
Before summing over all possible partitions, let us apply the remaining longitudinal operator $\mathcal{L}_x$, which will act only on the matrix elements $C_{jx}$ in \eqref{longitudinal_odd}. The longitudinal operator acts on a diagonal element of $C$ as
\begin{equation}
\mathcal{L}_{x} C_{xx} = \mathcal{L}_x \cdot \sum_{c \neq x}\frac{e_x p_c}{\sigma_{xc}} = \sum_{c \neq x}\frac{p_x p_c}{\sigma_{xc}} = 0
\end{equation}
on the support of the scattering equations. Therefore, the action of $\mathcal{L}_x$ on the matrix $C$ is given by
\begin{equation}
\mathcal{L}_x C_{jx} = \begin{cases} \frac{p_j p_x}{\sigma_{jx}} & \text{if } j \neq x \\ 0 & \text{if } j=x\end{cases} \,.
\end{equation} 
Since we can always choose $p$ and $q$ such that they are not equal to $x$, we have 
\begin{equation}
\begin{split}
\sum_{\substack{j=1\\j \neq p,q}}^n (-1)^{j+1} C_{jx} \mathrm{Pf}'(A_j^j) \xrightarrow{\;\;\mathcal{L}_x\;\;} \sum_{j \neq p,q,x} (-1)^{j+1} A_{jx} \mathrm{Pf}'(A_j^j) = \mathrm{Pf}'(\bar{A})
\end{split}
\end{equation}

\noindent
For the equal sign the recursion relation of the Pfaffian was used backwards to obtain a new $(n+1) \times (n+1)$ matrix $\bar{A}$. This matrix is an extension of the matrix $A$ and results from duplicating the x-th row and x-th column and adding them to the end. 
Therefore, since the rows (and columns) of $\bar{A}$ are not linearly independent, $\mathrm{Pf}'(\bar{A}) = 0$  for $n$ odd:
\begin{equation}
\label{n_odd}
\mathcal{T}_{a_1a_2} \cdot \prod_{k=1}^{n-2} \mathcal{L}_{b_k} \cdot \mathrm{Pf}'(\Psi) = 0  \,.
\end{equation}
\end{itemize}

\noindent
Using the fact that the Pfaffian of an antisymmetric $n \times n$ matrix vanishes for $n$ odd, the results for both cases can be summarized as
\begin{equation}
\mathcal{T}_{a_1a_2} \cdot \prod_{k=1}^{n-2} \mathcal{L}_{b_k} \cdot \mathrm{Pf}'(\Psi) = (-1)^\frac{n}{2}\mathrm{Pf}'(A)^2 \,.
\end{equation}

We can now act with the operator \eqref{longitudinal_op} on the CHY integrand of gravity to obtain the BI integrand containing $n$ BI photons. Due to the permutation invariance of the reduced Pfaffian, this integrand is independent of the choice of the two trace-operator transmuted particles
\begin{equation}
\begin{split}
\mathcal{I}_{\mathrm{BI}}(\tilde{\gamma}_1,\dots,\tilde{\gamma}_n) = &\; \mathcal{T}_{a_1a_2} \cdot \prod_{k=1}^{n-2} \mathcal{L}_{b_k} \cdot \mathcal{I}_{\mathrm{G}}(h_1,\dots,h_n) 
 = \;(-1)^\frac{n}{2}\,\left[ \mathrm{Pf}'A(p,\sigma)\right]^2 \, \mathrm{Pf}'\Psi(p,\tilde{e},\sigma) \,.
\end{split}
\end{equation}
The notation $\tilde{\gamma}_i$ denotes that the BI photon $i$ is characterized by the polarization vector $\tilde{e}_i$. Applying the operator once more to $\mathrm{Pf}'\Psi$  leads to the SG integrand  with $n$ SG scalars: 
\begin{equation}
\mathcal{I}_{\mathrm{SG}}(\phi_1,\dots,\phi_n) = \mathcal{T}_{a_1a_2} \cdot \prod_{k=1}^{n-2} \mathcal{L}_{b_k} \cdot \mathcal{I}_{\mathrm{BI}}(\gamma_1,\dots,\gamma_n)  = \left[ \mathrm{Pf}'A(p,\sigma) \right]^4 \,.
\end{equation}
By proceeding in a similar way, the  integrand for pure NLSM of $n$ pions is obtained by applying the $\lvert \alpha \rvert = 2$ longitudinal operator to the YM integrand 
\begin{equation}
\begin{split}
\mathcal{I}_{\mathrm{NLSM}}^{[\omega]}(\pi_1,\dots,\pi_n) = &\; \mathcal{T}_{a_1a_2} \cdot \prod_{k=1}^{n-2} \mathcal{L}_{b_k} \cdot \mathcal{I}_{\mathrm{YM}}^{[\omega]}(g_1,\dots,g_n)\\[1em]
  =&\; \frac{-[\mathrm{Pf}'A(k,\sigma)]^2}{\sigma_{\omega(1),\omega(2)} \cdots \sigma_{\omega(n),\omega(1)}} \,.
\end{split}
\end{equation}

The NLSM integrand  can also be obtained by applying a $\mathcal{T}[\omega]$ operator to the integrand of BI theory: this is the same procedure above but with operators in reversed order. Since the transmutation operators commute among themselves, the order of application does not matter.
With the same operation, we could also retrieve the DBI integrand from the EM one.

These results agree with \cite{Cachazo:2014xea} up to an irrelevant sign. Let us notice that the trace operator $\mathcal{T}_{a_1a_2}$ in equation \eqref{longitudinal_op} guarantees that the reduced Pfaffian of $A$ is taken instead of the full Pfaffian, which would be the result with only longitudinal operators. This is the operatorial equivalent of the conjectured procedure described in \cite{Cachazo:2014xea}.

%%%%%%%%%%%%%%

Finally, we investigate the extended theories generated by the operator $\mathcal{T}[\alpha]\mathcal{L}^{\bar{\alpha}}$ defined in \eqref{longitudinal1}, with $\lvert \alpha \rvert > 2$. The set of all particles $O_n$ is split into two subsets $\alpha = \{ a_1,a_2,\dots,a_m \}$ and $\bar{\alpha} = \{ a_{m+1},a_{m+2},\dots,a_n \}$, where all particles in $\alpha$ ($\bar{\alpha}$) are transmuted according to $\mathcal{T}$ ($\mathcal{L}$). In the case where $\lvert\bar{\alpha}  \rvert$ is odd, the operator in \eqref{longitudinal1} leads to a vanishing result, as we already showed in (\ref{n_odd}) for the particular case $\lvert \alpha \rvert = 2$. 
 Using our previous results, we can easily study the action of $\mathcal{T}[\alpha]\mathcal{L}^{\bar{\alpha}}$ on the reduced Pfaffian $\mathrm{Pf}'\Psi$. The action of the multiple trace operator is governed by \eqref{gen_st}
\begin{equation}
\mathcal{T}[\alpha] \cdot \mathrm{Pf}'(\Psi) = s_{m,n} \; C_m [\alpha] \; \mathrm{Pf}[\Psi]_{\bar\alpha;\bar\alpha}.
\end{equation}
The second part $\mathcal{L}^{\bar{\alpha}}$ acts now on the Pfaffian of the $2(n-m){\times}2(n-m)$ matrix $\mathrm{Pf}[\Psi]_{\bar\alpha;\bar\alpha}$ of the remaining particles belonging to $\bar{\alpha}$. 
The $n-m$ one-point longitudinal operators can again be expanded in terms of two-point longitudinal operators.
Studying one particular partition, we find
\begin{equation}
\prod_{k=1}^r \mathcal{L}_{i_kj_k} \cdot \mathrm{Pf}[\Psi]_{\bar\alpha;\bar\alpha} = \mathrm{sgn}(\{ ij\}) \; \prod_{k=1}^r \left( -\frac{p_{i_k}p_{j_k}}{\sigma_{i_kj_k}} \right) \; \mathrm{Pf} [A]_{\bar\alpha;\bar\alpha},
\end{equation}
where  $r=\frac{n-m}{2}$ and $\mathrm{Pf}[A]_{\bar\alpha;\bar\alpha}$ is the matrix $A$ restricted to particles from the set $\bar{\alpha}$. Summing over all partitions, the terms combine into another Pfaffian of the matrix $[A]_{\bar\alpha;\bar\alpha}$:
\begin{equation}
\begin{split}
\mathcal{L}^{\bar{\alpha}} \cdot \mathrm{Pf}[\Psi]_{\bar\alpha;\bar\alpha} = & (-1)^r \; \mathrm{Pf} [A]_{\bar\alpha;\bar\alpha} \; \sum_{ \{ij\} } \mathrm{sign}(\{ ij\}) \; \prod_{k=1}^r \left( A_{i_kj_k} \right) =  -s_{n-m,n-m} \left( \mathrm{Pf}[A]_{\bar\alpha;\bar\alpha}\right)^2.
\end{split}
\end{equation}
Therefore, the full operator gives
\begin{equation}\label{long:extended}
\mathcal{T}[\alpha]\mathcal{L}^{\bar{\alpha}} \cdot \mathrm{Pf}'(\Psi) = (-1)^{\frac{n+m^2}{2}-1} \; C_m [\alpha] \; \left( \mathrm{Pf}[A]_{\bar\alpha;\bar\alpha}\right)^2.
\end{equation}

Applying this operator to the integrand of gravity, all gravitons in the set $\alpha$ are transmuted into gluons ordered with respect to $\alpha$ and all gravitons in $\bar{\alpha}$ are transmuted into BI photons. The integrand of the resulting extended BI theory is given by
\begin{equation}\label{long:BIx}
\mathcal{I}_{\mathrm{BI}\oplus\mathrm{YM}} (\alpha) = \tilde{\mathcal{T}}[\alpha]\tilde{\mathcal{L}}^{\bar{\alpha}} \cdot \mathcal{I}_{\mathrm{G}} = (-1)^{\frac{n+m^2}{2}-1} \; C_m [\alpha] \; \left( \mathrm{Pf}[A]_{\bar\alpha;\bar\alpha}\right)^2 \; \mathrm{Pf}' \Psi.
\end{equation}
From here a second copy of the $\mathcal{T}[\beta]\mathcal{L}^{\bar{\beta}}$  operator leads to an extended SG theory which couples SG scalars and biadjoint scalars to pions from two copies of NLSM:
\begin{eqnarray}\label{long:SGx}
\mathcal{I}_{\mathrm{SG}\oplus\mathrm{NLSM}^2\oplus\mathrm{BS}} (\alpha | \beta) &=&  \mathcal{T}[\beta]\mathcal{L}^{\bar{\beta}} \cdot \mathcal{I}_{\mathrm{BI}} \nonumber\\
&=& (-1)^{\frac{m^2+m'^2}{2}+n} \; C_m[\alpha] \; \left( \mathrm{Pf}[A]_{\bar\alpha;\bar\alpha}\right)^2 \; C_{m'} [\beta] \; \left( \mathrm{Pf}[A]_{\bar\beta;\bar\beta}\right)^2
\end{eqnarray}
with $m' = \lvert \beta \rvert$. The particles are transmuted in four distinct groups depicted in the table below. The intersection $\alpha \cup \beta$ contains biadjoint scalars with orderings $\alpha$ and $\beta$ and the intersection $\bar{\alpha} \cup \bar{\beta}$ SG scalars. The particles in the two remaining intersections $\alpha \cup \bar{\beta}$ and $\bar{\alpha} \cup \beta$ are pions of two copies of NLSM, the first with color structure $\alpha$ and the other with color structure $\beta$.
\renewcommand{\arraystretch}{1}
\begin{center}
\begin{tabular}{| c | c | c |}
\hline
& $\alpha$ & $\bar{\alpha}$ \\\hline
$\beta$ & BS & NLSM \\\hline
$\bar{\beta}$ & NLSM & SG \\\hline
\end{tabular}
\end{center}
Applied to the $n$-gluon YM integrand with ordering $\alpha$, the generalized operator $\mathcal{T}[\beta]\mathcal{L}_{\bar{\beta}}$ transmutes all gluons in $\beta$ into biadjoint scalars and all gluons in $\bar{\beta}$ into NLSM pions. This leads to the following integrand of extended NLSM:
\begin{equation}\label{long:NLSMx}
\mathcal{I}_{\mathrm{NLSM}\oplus\mathrm{BS}} (\alpha | \beta) = \mathcal{T}[\beta]\mathcal{L}^{\bar{\beta}} \cdot \mathcal{I}_{\mathrm{YM}} = (-1)^{\frac{n^2+m^2}{2}+n} \; C_n[\alpha] \; C_m [\beta] \; \left( \mathrm{Pf}[A]_{\bar\beta;\bar\beta}\right)^2
\end{equation}
All particles are placed in one color trace ordered by $\alpha$ and the biadjoint scalars are additionally ordered with respect to their dual color with ordering $\beta$.

These results agree with the integrands of extended BI, SG and NLSM described in \cite{Cachazo:2016njl}.

%%%%%%%%%%%%%%%%%%%%%%%%%%%%%%%%%%%%
%%%%%%%%%%%%%%%%%%%%%%%%%%%%%%%%%%

\begin{figure}
\centering
\begin{tikzpicture}[scale=1]
\node[align=center, below, draw, rectangle, fill=blue!10](G) at (0,0){\underline{Gravity}\\$\mathrm{Pf}' \Psi$ \\ $\mathrm{Pf}' \tilde{\Psi}$};
\node[align=center, below, draw, rectangle](YM) at (4,0){\underline{YM}\\$C_n[\alpha]$ \\ $\mathrm{Pf}' \tilde{\Psi}$};
\node[align=center, below, draw, rectangle](BS) at (7.5,0){\underline{BS}\\$C_n[\alpha]$ \\ $C_n[\beta]$};
\node[align=center, below, draw, rectangle](BI) at (0,-3.5){\underline{BI}\\$\mathrm{Pf}'(A)^2$ \\ $\mathrm{Pf}' \tilde{\Psi}$};
\node[align=center, below, draw, rectangle](NLSM) at (4,-3.5){\underline{NLSM}\\$C_n[\alpha]$ \\ $\mathrm{Pf}'(A)^2$};
\node[align=center, below, draw, rectangle](SG) at (0,-6.5){\underline{SG}\\$\mathrm{Pf}'(A)^2$ \\ $\mathrm{Pf}'(A)^2$};
\node[align=center, below, draw, rectangle](EM) at (-3,3){\underline{EM}\\$\mathrm{Pf}' [\Psi]_{h,\gamma;h}$ \\ $\mathrm{Pf}[\mathcal{X}]_{\gamma} \, \mathrm{Pf}' \tilde{\Psi}$};
\node[align=center, below, draw, rectangle](YMS) at (1,3){\underline{YMS}\\$C_n[\alpha]\, \mathrm{Pf}[\mathcal{X}]_{s}$ \\ $\mathrm{Pf}' [\Psi]_{g,s;g}$};
\node[align=center, below, draw, rectangle](DBI) at (-3,-0.5){\underline{DBI}\\$\mathrm{Pf}'(A)^2 \, \mathrm{Pf}[\mathcal{X}]_{s}$\\$\mathrm{Pf}' [\Psi]_{\gamma,s;\gamma}$};
\draw[->] (G) -- (YM) node[pos=0.5,anchor=south]{\textcolor{rgb:red,3;black,1}{$\mathcal{T}[\alpha]$}};
\draw[->] (YM) -- (BS) node[pos=0.5,anchor=south]{\textcolor{rgb:red,3;black,1}{$\tilde{\mathcal{T}}[\beta]$}};
\draw[->,densely dotted] (G) -- (BI) node[pos=0.5,anchor=west]{\textcolor{rgb:blue,5;black,1}{$\mathcal{T}\mathcal{L}$}};
\draw[->] (BI) -- (NLSM) node[pos=0.5,anchor=south]{\textcolor{rgb:red,3;black,1}{$\tilde{\mathcal{T}}[\alpha]$}};
\draw[->,densely dotted] (YM) -- (NLSM) node[pos=0.5,anchor=west]{\textcolor{rgb:blue,5;black,1}{$\tilde{\mathcal{T}}\tilde{\mathcal{L}}$}};
\draw[->,densely dotted] (BI) -- (SG) node[pos=0.5,anchor=west]{\textcolor{rgb:blue,5;black,1}{$\tilde{\mathcal{T}}\tilde{\mathcal{L}}$}};
\draw[->,dashed] (G) -- (EM) node[pos=0.5,anchor=west]{\textcolor{rgb:green,1;black,1}{$\mathcal{T}_{\gamma}^m$}};
\draw[->,dashed] (YM) -- (YMS) node[pos=0.5,anchor=west]{\textcolor{rgb:green,1;black,1}{$\tilde{\mathcal{T}}_{s}^m$}};
\draw[->] (EM) -- (YMS) node[pos=0.5,anchor=south]{\textcolor{rgb:red,3;black,1}{$\tilde{\mathcal{T}}[\alpha]$}};
\draw[->,dashed] (BI) -- (DBI) node[pos=0.5,anchor=west]{\textcolor{rgb:green,1;black,1}{$\tilde{\mathcal{T}}_{s}^m$}};
\draw[->,densely dotted] (EM) -- (DBI) node[pos=0.5,anchor=west]{\textcolor{rgb:blue,5;black,1}{$\tilde{\mathcal{T}}\tilde{\mathcal{L}}$}};
\end{tikzpicture}
\caption{Summary of the action of transmutation operators on CHY integrands.  Denoting the set of gravitons by $h$,
the set of gluons by $g$, the set of photons by $\gamma$ and the set of scalars by $s$, we write $[\mathcal{X}]_x$
for the matrix $\mathcal{X}$ restricted to the specific subset $x$ and $[\Psi]_{x,y;x}$ for the matrix $\Psi$ with
its full first part and its second part restricted to $x$. All integrands are presented modulo
an overall sign.} \label{fig:connections}
\end{figure}
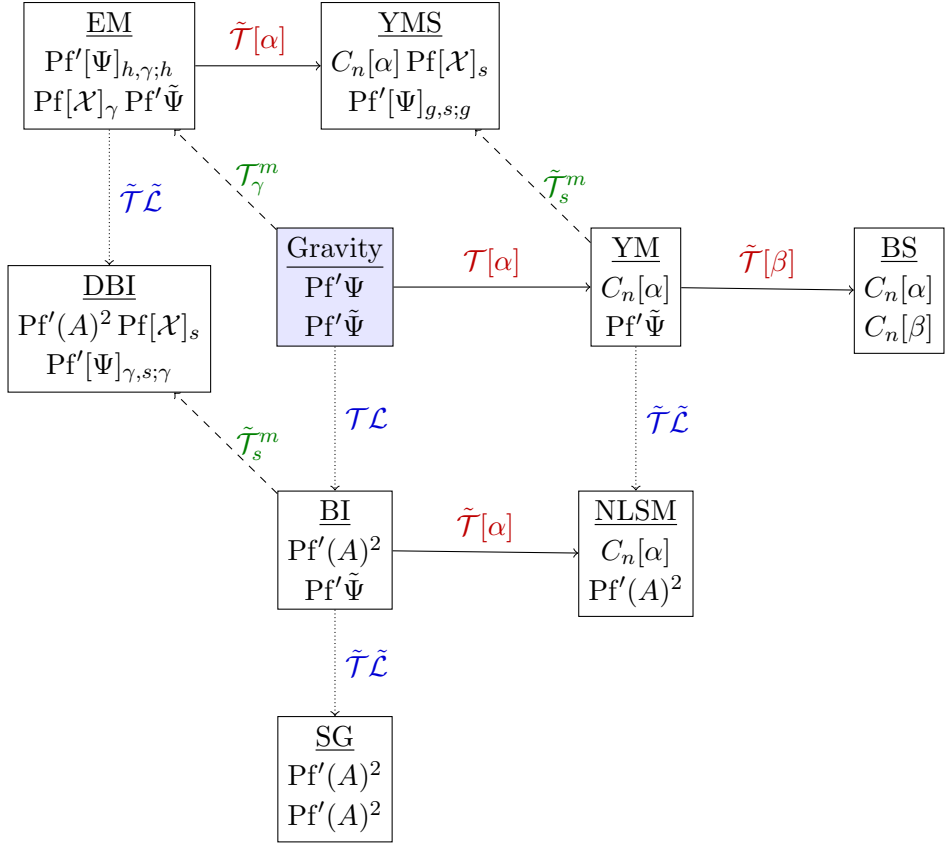
%%%%%%%%%%%%%%%%%%%%%%%%%%%%%%%%%%%%
%%%%%%%%%%%%%%%%%%%%%%%%%%%%%%%%%%

\section{A first look at loop amplitudes}
\label{one-loop}

By using ambitwistor string, in \cite{Geyer:2017ela} was shown that the one-loop integrands of $n$-point scattering amplitudes for gravity and Yang-Mills exhibit a very similar structure to those at tree-level, i.e. a universal measure depending on the so-called nodal scattering equations and integrands built from few building blocks:
\begin{equation}
\label{CHY_1loop}
A_n^{(1)} = \int\frac{d^D\ell}{\ell^2}   \int_{\mathfrak{M}_{0,n+2}}  \frac{\prod_a \mathrm{d}\sigma_a}{\mathrm{vol}\,\mathrm{SL}\left(2,\mathbb{C}\right)} \,\prod_{a}~' \delta \left( E_a^{\mathrm{(nod)}} \right) \, \mathcal{I}^{(1)}_{n}(\{p, e, \sigma\})  =:  \int\frac{d^D\ell}{\ell^2} \int_{\mathfrak{M}_{0,n+2}} d\mu_{1,n} \, \mathcal{I}^{(1)}_{n}  \,,
\end{equation}
where $a= 1, \cdots,n, +, -$ and
\begin{equation}
E_+^{\mathrm{(nod)}}=\sum_{i}\frac{\ell k_i}{\sigma_{+i}} \quad ,\quad E_-^{\mathrm{(nod)}}=-\sum_{i}\frac{\ell k_i}{\sigma_{-i}} \quad ,\quad E_i^{\mathrm{(nod)}}=   \frac{\ell k_i}{\sigma_{i+}}- \frac{\ell k_i}{\sigma_{i-}}+\sum_{j\neq i}\frac{k_i k_j}{\sigma_{ij}}
\end{equation}
are the $n+2$ nodal scattering equations. The one-loop measure $\mathrm{d}\mu_{1,n}$ is the same as the tree-level one, $\mathrm{d}\mu_{0,n+2}$, with two additional particles carrying momentum $\pm \tilde{\ell}$ and labeled by the punctures $\sigma_+$ and $\sigma_-$ respectively.
The on-shell momentum $\tilde{\ell}$ is related to the loop momentum $\ell$ via $\tilde{\ell} = \ell + \eta$ with an auxiliary momentum $\eta$ satisfying $\ell \eta = p_i\eta = e_i\eta = 0$. 
Also the integrand $\mathcal{I}_n^{(1)}$ can be related to the tree-level integrand $\mathcal{I}_{n+2}^{(0)}$ .
The same structure was found  for biadjoint scalar amplitudes in  \cite{He:2015yua}  using the forward limit.

The building blocks of the one-loop integrands are given by the $(n\! + \! 2)$-particle Parke-Taylor factor 
\begin{equation}
C_{n+2}[\alpha] = \frac{1}{\sigma_{+\alpha(1)} \sigma_{\alpha(1)\alpha(2)} \cdots \sigma_{\alpha(n)-} \sigma_{-+}}
\end{equation}
and the one-loop NS integrand $\mathcal{I}_{\mathrm{NS}}^{(1)}$ replacing the reduced Pfaffian $\mathrm{Pf}' \Psi$. The NS integrand reads 
\begin{equation}
\label{INS}
\mathcal{I}_{\mathrm{NS}}^{(1)} = \sum_r \mathrm{Pf}' \left( \Psi_{\mathrm{NS}}^r \right),
\end{equation}
where the matrix $\Psi_{\mathrm{NS}}^r$ is given by extending the tree-level matrix $\Psi$ defined in equation \eqref{PsiMatrix} to $n\! + \! 2$ points with the two additional particles carrying momentum $\pm \tilde{\ell}$ and polarizations $e_+ = e^r$  and $e_- = \left( e^r \right)^{\dagger}$ respectively:
\begin{equation}
\Psi_{\mathrm{NS}}^r = \Psi^{\mathrm{tree}}_{n+2}\Big{|}_{e_+ = e^r,  e_- = \left( e^r \right)^{\dagger},\, \tilde{l}^2=0} \,.
\end{equation}
Analogously to the tree-level case, the one-loop integrands for gravity, color-ordered Yang-Mills and biadjoint scalar are given by
\begin{eqnarray}
\mathcal{I}_{\mathrm{G}}^{(1)} &=& \mathcal{I}_{\mathrm{NS}}^{(1)} \; \tilde{\mathcal{I}}_{\mathrm{NS}}^{(1)}\\
\mathcal{I}_{\mathrm{YM}}^{(1)}(\alpha) &=& C_{n+2}[\alpha] \; \mathcal{I}_{\mathrm{NS}}^{(1)}\\
\mathcal{I}_{\mathrm{BS}}^{(1)}(\alpha|\beta) &=& C_{n+2}[\alpha] \; C_{n+2}[\beta] \,.
\end{eqnarray}

Therefore, the one-loop CHY integrands are connected in the same way by transmutation operators as the tree-level ones. In this case the transmutation operators have to be extended to $n + 2$ points, including the two particles $+$ and $-$, and have to be applied before summing over the internal degrees of freedom $r$. In this way, we can relate the integrands above  via the operator $\mathcal{T}[12 \cdots n - +] $ by applying it to the reduced Pfaffian of the $2(n+2) \times 2(n+2)$ matrix $ \Psi^r_{NS} $, before taking the sum over $r$ in equation \eqref{INS}. Analogously to the tree-level case, the result is the $(n+2)$-point Parke-Taylor factor:
\begin{equation}
\mathcal{T}[12 \cdots n - +]  \Psi^r_{\mathrm{NS}} = C_{n+2}[12 \cdots n  - +] \,,
\end{equation}
leading to the relations
\begin{equation}
\mathcal{I}_{\mathrm{G}}^{(1)} \rightarrow  \mathcal{I}_{\mathrm{YM}}^{(1)}(\alpha) \rightarrow \mathcal{I}_{\mathrm{BS}}^{(1)}(\alpha|\beta) \,.
\end{equation}

Following the same steps as in the previous sections, one can derive the one-loop integrands for all the other  theories and their extensions and find similar results as those collected in Figure \ref{fig:connections}. 
The similarity between tree-level and one-loop amplitudes finds its reason in the Feynman tree theorem and it is realised in the ambitwistor string context by the so-called gluing operator \cite{Roehrig:2017gbt}. This indeed relates amplitudes at one loop to tree-level amplitudes with two additional particles. However, this theorem does not hold at two or more loops. Therefore, we do not expect the integrands at two or more loops to be connected by the transmutation operators  in the same way.

%%%%%%%%%%%%%%%%%%%%%%%%%%%%%%%%%%%%
%%%%%%%%%%%%%%%%%%%%%%%%%%%%%%%%%%

\section{Conclusions and Outlook}

In this paper we applied the recently proposed transmutation operators \cite{Cheung:2017ems} to the CHY formalism of scattering amplitudes and reconstructed CHY integrands for a wide range of theories. Starting from gravity, we obtained integrands for Yang-Mills, biadjoint scalar, Einstein-Maxwell, Yang-Mills scalar, Born-Infeld, Dirac-Born-Infeld, non-linear sigma model and special Galileon theory. The expressions of these integrands are shown in Figure \ref{fig:connections} and agree with the known integrands of these theories \cite{Cachazo:2013iea, Cachazo:2014xea}. This verifies the form of some conjectured CHY integrands for any number of points.  
Moreover, we derived the CHY integrands for the extensions of BI, NLSM and SG by linking single-trace and longitudinal operators. These integrands from extended theories agree with the results described in \cite{Cachazo:2016njl}. The computations in chapter \ref{transm_CHY} provide a powerful tool for computing integrands for arbitrary theories by applying the appropriate combination of transmutation operators.
By using the similarity among one-loop $n+2$-points and tree-level $n$-points integrands, we  derived the one-loop CHY integrands for various theories by extending the transmutation operators to $n+2$ points. This is done in a similar way as the tree-level case. However, we do not expect the integrands at two or more loops to be connected via the transmutation operators  \cite{Cheung:2017ems} in the same way. Indeed, at  higher loops, the integrands cannot be constructed from the tree-level ones and a suitable modification of these operators has yet to be found.
We leave this interesting question for future work.

%%%%%%%%%%%%%%%%%%%%%%%%%%%%%%%%
%%%%%%%%%%%%%%%%%%%%%%%%%%%%%%%

\section*{Acknowledgements}

We would like to thank Yvonne Geyer, Tomasz \L ukowski and Matteo Parisi for useful discussions. L.F. is supported by the Elitenetwork of Bavaria. This work was partially supported by the DFG Grant FE 1529/1-1.

%%%%%%%%%%%%%%%%%%%%%%%%%%%%%%%%%%%%%%%%%%%%%%
%%%%%%%%%%%%%%%%%%%%%%%%%%%%%%%%%%%%%%%%%%%%%%

\appendix
\addtocontents{toc}{\protect\setcounter{tocdepth}{0}}

\section{The Pfaffian}
\label{app:pfaffian}

Let $M$ be an antisymmetric $2n{\times}2n$ matrix whose elements are denoted by $m_{ij}$. Then the Pfaffian of $M$ is defined as
\begin{equation}\label{pfaffian_def}
\mathrm{Pf}(M) = \sum_{\rho \in \Pi} \mathrm{sgn}(\pi_{\rho}) \prod_{k=1}^n m_{i_kj_k}
\end{equation}
where $\Pi$ is the set of all partitions of $\{ 1,2,...,2n\}$ into pairs without regard to order. A generic element $\rho \in \Pi$ can be represented as an ordered set of $n$ ordered pairs
\begin{equation}
\rho = \{ (i_1,j_1),...,(i_{n},j_{n}) \} \hspace{0.5cm} \mathrm{with} \; i_k < j_k \; \mathrm{and} \; i_1 < i_2 < ... < i_{n}.
\end{equation}
To every $\rho \in \Pi$ one associates a permutation $\pi_{\rho}$ with signature $\mathrm{sgn}(\pi_{\rho})$
\begin{equation}
\pi_{\rho} = \begin{pmatrix}
1 & 2 & \dots & 2n-1 & 2n \\
i_1 & j_1 & \dots & i_n & j_m
\end{pmatrix}.
\end{equation}
In other words one first builds a product of components $m_{ij}$ such that every index appears only once and $i<j$ is true for every component. Such a product will consist of $n$ factors. In a second step one sums over all $(2n-1)!!$ possible (and different) products that can be obtained in this way, where every even permutation comes with a plus sign in front and every odd permutation with a minus sign.

The Pfaffian fulfills the following recursion relation:
\begin{equation}\label{recursivePfaffian}
\mathrm{Pf}(M) = \sum_{\substack{j=1\\j\neq i}}^{2n}(-1)^{i+j+1+\Theta(i-j)} \, m_{ij} \, \mathrm{Pf}(M_{i,j}^{i,j})
\end{equation}
for $i \in \{1,2,...,2n \}$. Here $\Theta$ denotes the Heaviside step function and $M_{i,j}^{i,j}$ the matrix $M$ with rows and columns $i$ and $j$ removed. Generalizing this notation we will write $M_{i_1,...,i_m}^{i_1,...,i_m}$ for the matrix $M$ with rows and columns $i_1,...,i_m$ removed.

%%%%%%%%%%%%%%%%%%%%%%%%%%%%%%%%%%%%%%%%%%%%%%%%%%%
%%%%%%%%%%%%%%%%%%%%%%%%%%%%%%%%%%%%%%%%%%%%%%%%%%%

\bibliographystyle{nb}
\bibliography{Transms}

\end{document}